# Evolution of vortices created by conical diffraction in biaxial crystals versus orbital angular momentum

A. Brenier

Univ Lyon, Université Claude Bernard Lyon 1, CNRS, Institut Lumière Matière, F-69622, LYON, France

Abstract: Light states evolution versus their fractional orbital angular momentum (OAM) has been analyzed in the conical diffraction process occurring through biaxial crystals. Experimental results are provided by a non-degenerate cascade of $KGd(WO_2)_4$ and $Bi_2ZnOB_2O_6$ biaxial crystals. The continuous 0→1→2 $\hbar$/photon increasing of the fractional OAM in passing through integer values was operated with the help of the spin-orbit coupling in the $Bi_2ZnOB_2O_6$ crystal. The phase of the state light and its vortices were visualized by interference patterns with a reference beam. The evolution of the fractional OAM value is accompanied by a continuous evolution of pairs of vortices with opposite signs and linked by a -π/+π discontinuous phase line. The phase pattern evolution around half-integer OAM is observed to be continuous. In other cases, the evolution can be interrupted by the breaking of a -π/+π discontinuous phase line and a new pair of vortices with opposite charges is born.

1. Introduction

A beam going through a material is refracted and even generally doubly-refracted if the material is birefringent. This very common behaviour does not occur in the singular case of propagation along the optical axis of a biaxial crystal: W. Hamilton predicted in the nineteenth century that the refraction should be conical and a hollow cylinder should emerge [1, 2]. However, the whole optical phenomenon is quite complex and cannot be reduced to a simple propagation of rays. A more brilliant theoretical description was obtained from plane waves decomposition [3] of the incident beam. Their recombination behind the crystal reconstitutes the conical diffraction [4] of the output beam and the richness of its structure: internal and external conical refraction both included, light dual-cone, Raman spikes and Poggendorff rings [5]. Cascading several biaxial crystals leads to multiple rings and useful mode conversion. The theory for one crystal was generalised to N cascaded crystals [6]. The intensity patterns of multiple rings obtained with up to three identical crystals were experimentally exhibited [7]. In Ref. [8] an elliptical beam was launched into two cascaded crystals, reducing the intensity pattern of each circle in two lobes. Much more complex intensity patterns can be obtained inserting optical elements between the crystals. This was shown experimentally and theoretically in Ref. [9] with a λ/4 or a λ/2 plate or a linear polarizer inserted between two or more crystals. With the help of polarizers, $LG_0^1$, $LG_1^1$ and $LG_0^2$ Laguerre-Gauss were exhibited in Ref. [10] with two cascaded crystals.

Interestingly, Berry et al. [11] showed that the beam emerges also with a modified orbital angular momentum (OAM). Let us recall that the OAM results of the beam linear momentum acting off-axis with respect to its centre [12]. For example, Laguerre-Gauss modes have such $l\,\hbar$/photon OAM related to their $\exp(il\varphi)$ transverse phase variation. For an input field with a circular polarization, the output field after CD is a superposition of a $B_0$ component with a nil OAM (charge 0 component) and a $B_1$ component with 1 $\hbar$/photon OAM (charge 1 component). These predictions were experimentally verified in the case of the centrosymmetric $KGd(WO_2)_4$ (KGW) crystal [13] and the non-centrosymmetric (Pasteur medium) $Bi_2ZnOB_2O_6$ (BZBO) oxy-borate crystal [14].

The coherent superposition of several light states with integer OAM leads to fields with average fractional OAM. This situation is mainly documented with beams emerging from spiral phase plates with fractional step height, a spatial light modulator [15-17] and superposition of appropriate LG modes with different Gouy phases [18]. The case of cascaded biaxial crystals is mainly documented for integer OAM [19-20].

The present work is devoted to the evolution of light states versus their fractional OAM, generated by cascading two different biaxial crystals optical axis oriented: KGW and BZBO (non-degenerate cascade). The phase and intensity of the coherent superposition of modes, their visualization by interference fringes and CCD camera, and their theoretical calculations by a full numerical model are presented versus the fractional OAM, including around half-integer values.

## 2. Experimental methods

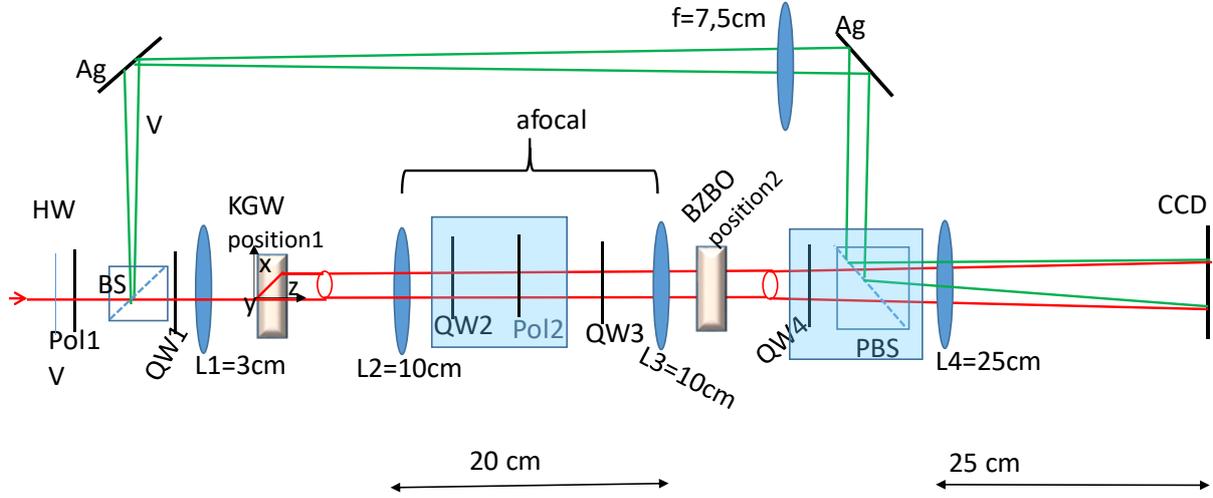

Fig. 1 Experimental set-up allowing the OAM study in the far field. The CD-field in position 1 is transported without modification in position 2 by the L2-L3 afocal telescope.

A KGW (6 X 6X 3 mm$^3$) and a BZBO (8 X6 X 6 mm$^3$) crystals were used, both oriented along their optical axis. Their x-z principal plane was kept horizontal.

The linear polarized beam from a 1064 nm YAG:Nd laser was first made left circular (LC) polarized with a quarter-wave plate at the entrance of the set-up (Fig. 1). It was focused through the KGW into a 40 μm waist spot a few mm behind the sample with a doublet lens L1. In the focal plane labelled "position 1" in Fig. 1, we can observe the ring of the CD located between two axial spikes. In order to have enough space to locate several optical elements, the CD phenomenon in "position 1" was reproduced exactly in amplitude, phase and structure (without magnification) 40 cm further in "position 2" with an afocal telescope constituted from two 10 cm focal length doublets L2 and L3, separated by twice (20 cm) their focal length. The BZBO biaxial sample was located in this position and it modified the CD phenomenon. The far field resulting from the cascaded CD and propagating further position 2 was visualized through a Fourier lens L4 on the screen of a Beamage Gentec CCD camera located in the L4 focal plane. In order to visualise the phase of the near field, a reference beam was taken from the input beam with a 50/50 beam splitter and redirected with mirrors towards the CCD screen by a polarizing beam splitter near "position 2" and through the Fourier lens. The whole reference set-up is similar to a Mach-Zehnder interferometer. On the output face of the beam-splitter the two beams were a few mm separated in order to create a controlled wedge fringe pattern with the desired spatial resolution. The polarizing beam splitter reflects the vertically polarized reference beam and transmits the horizontal component of the beam under study, so their interference on the CCD is obtained through a supplementary rotated linear polarizer not shown in Fig. 1.

In Fig. 1 we can see several optical elements inserted inside the beam path: linear polarizers and quarter-wave plates, used for selecting the light states as it is detailed further.

## 3. Theoretical background

### 3.1 Beam propagation through a biaxial crystal in the Fourier space and in a circular basis

The electric field **E** of a monochromatic beam at the input of the crystal can be decomposed in plane waves. Each of them wave inside the crystal has components on the two eigen-modes (±) of the refracted propagation direction and it propagates as $\exp[i(k_x x + k_y y + k_{z\pm} z)]$, with:

$$k_{z\pm} = \sqrt{k_\pm^2 - (k_x^2 + k_y^2)} \qquad (1)$$

$k_x, k_y$ being the transverse components of the wave-vector and $k_\pm$ being the two eigen-values. At the exit of the crystal with thickness $L$ the emerging field $\left[\hat{\mathbf{E}}(L, k_x, k_y)\right]_{xy}$ is obtained from a linear operator $\hat{U}$ acting on $\left[\hat{\mathbf{E}}(0, k_x, k_y)\right]_{xy}$:

$$\hat{U} = exp\left\{izP \begin{bmatrix} k_{z+} & 0 \\ 0 & k_{z-} \end{bmatrix} P^{-1}\right\} \qquad (2)$$

where $P$ is the transfer matrix from the xy-frame towards the eigen-modes frame. The parabolic approximation leads to smart results for the eigen-values, the eigen-modes and the operator $\hat{U}$ [3, 4]. In the present work we operated a full numerical calculation (with the Matlab package) of the $\hat{U}$ operator and of any fields such as $\widehat{\mathfrak{L}}$ and $\widehat{\mathfrak{R}}$ hereafter, with the exact refractive indices and eigenmodes as it is detailed in Ref. [14] and including birefringence and bi-anisotropy. Moreover, for any path of length $\Delta z$ in free space after the exit face of one crystal, it is necessary to multiply the field by $\exp(i\sqrt{k_0^2 - (k_x^2 + k_y^2)}\Delta z)$, $k_0$ being the free-space wave-vector. Of course the field in a given plane of the real space (near field, exhibited for example in Ref. [14]) is provided by the inverse Fourier transform, but in the present work we do not use this route, we work in the Fourier space (far field detected in the focal plane of the L4 Fourier lens). As it was recognized many years ago [11] waves with circular polarizations (CP) have a special role. Launching such a wave with no OAM inside a birefringent KGW crystal leads to two superimposed emerging waves $\mathfrak{L}$ (LCP) and $\mathfrak{R}$ (RCP) after CD in a birefringent biaxial crystal, the one with the opposite CP having its OAM modified as ±1 ℏ/photon, the one with the same CP having no OAM. The corresponding far fields $\widehat{\mathfrak{L}}$ and $\widehat{\mathfrak{R}}$ are constituted of concentric rings (Bessel beams), they are well studied generally from reasonable approximations applied to the refractive indices and propagation eigenmodes. So, they are known to be pure states of the z-component of the OAM operator:

$$\hat{L}_z = -i\hbar(k_x \frac{\partial}{\partial k_y} - k_y \frac{\partial}{\partial k_x}) \quad (3)$$

(Darwin result [21] restricted to the 2D monochromatic case). In a circular basis the CD through the KGW crystal labelled $i=1$ is obtained with the operator:

$$\widehat{U}_i(L, k_x, k_y) = \begin{bmatrix} \widehat{\mathfrak{L}}_i^{(0)} & \widehat{\mathfrak{L}}_i^{(-1)} \\ \widehat{\mathfrak{R}}_i^{(1)} & \widehat{\mathfrak{R}}_i^{(0)} \end{bmatrix} \quad (4)$$

where the superscript (c) is the OAM ($\hat{L}_z$ eigenvalue) of the CP wave. Let us add that our numerical calculations lead to 0.997 ℏ/photon instead of 1 ℏ/photon, meaning that our numerical error is about 0.3 %.

The case of the BZBO crystal, labelled $i=2$, needs a supplementary explanation. This is because this crystal is orthorhombic and acentric with point group mm2 ($C_{2v}$). It exhibits a natural optical activity, described by a tensor $\chi_a$, and responsible for the rotation of a linear polarization launched along the optical axis [14]. At 1064 nm, we measured the specific rotatory power $\rho$ to be 0.867 rad/cm. Strictly speaking, the far fields such $\widehat{\mathfrak{R}}_2^{(c)}$ are no more pure $\hat{L}_z$ eigenstates with integer OAM. Evaluating the c-charge of a $\begin{bmatrix} \widehat{\mathfrak{L}} \\ \widehat{\mathfrak{R}} \end{bmatrix}$ general field in the Fourier space with the Darwin formula:

$$l_z = \frac{\mathcal{R}_e \iint [\widehat{\mathfrak{L}}^*(\hat{L}_z\widehat{\mathfrak{L}}) + \widehat{\mathfrak{R}}^*(\hat{L}_z\widehat{\mathfrak{R}})] dk_x dk_y}{\iint [|\widehat{\mathfrak{L}}|^2 + |\widehat{\mathfrak{R}}|^2] dk_x dk_y} \quad (5)$$

where $\mathcal{R}_e$ means "real part", we found c=0.995 ℏ/photon. This is very slightly smaller than 1, in agreement with Berry prediction about the influence of chirality on CD. But taking into account the numerical uncertainty and the closeness of c to 1, we will neglect at this step the OAM mixture and we will used operator (4) with $i=2$.

Finally, in the following sections we will calculate the spin (ℏ/photon) of the field according to Darwin:

$$S_z = \frac{\iint \left[|\widehat{\mathfrak{L}}|^2 - |\widehat{\mathfrak{R}}|^2\right] dk_x dk_y}{\iint \left[|\widehat{\mathfrak{L}}|^2 + |\widehat{\mathfrak{R}}|^2\right] dk_x dk_y} \qquad (6)$$

**3.2 Beam with fractional orbital angular momentum**

In a previous work [20] the states with increasing integer OAM (0, 1, 2) after the BC2 crystal were studied in the real space (intensity and phase). At the opposite in the present work we focus on the states with fractional OAM in the Fourier space. Their evolution is studied from the $\theta$ angle continuously varied in the [-45°, +45°] interval.

The optical elements located after one crystal number *i* modify the wave polarization according to their Jones matrix in the circular basis (obtained with the help of the transfer matrix: $\frac{1}{\sqrt{2}}\begin{pmatrix} 1 & -i \\ 1 & i \end{pmatrix}$).

We can see in Fig. 1 two composite elements constituted by a quarter-wave plate with its slow axis making a $\alpha = \pm 45°$ angle with the x-horizontal axis, followed by a horizontal linear polarizer. They are shown inside two boxes (QW2-Pol2 and QW4-Pol3 with a polarizing beam splitter as pol3) because they are used as a whole. The Jones matrix of the box is:

$$J_\mathfrak{R} = \frac{1}{\sqrt{2}}\begin{bmatrix} 0 & 1 \\ 0 & 1 \end{bmatrix} \qquad (7)$$

if $\alpha = +45°$ and:

$$J_\mathfrak{L} = \frac{1}{\sqrt{2}}\begin{bmatrix} 1 & 0 \\ 1 & 0 \end{bmatrix} \qquad (8)$$

if $\alpha = -45°$ ($\alpha$ is respectively $\alpha_2$ or $\alpha_4$ for QW2-Pol2 box and QW4-Pol3 box).

The role of $J_\mathfrak{R}$ and $J_\mathfrak{L}$ in the experimental set-up is straightforward: $J_\mathfrak{R}$ projects a light state $\begin{bmatrix} \widehat{\mathfrak{L}}^{(cL)} \\ \widehat{\mathfrak{R}}^{(cR)} \end{bmatrix}$ onto a state with the charge *cR* of the right CP, while $J_\mathfrak{L}$ projects onto a state with the charge *cL* of the left CP. More, the two obtained states are horizontally polarized. In other words, the role of $J_\mathfrak{R}$ and $J_\mathfrak{L}$ is to select a wanted charge (the highest if the purpose is scaling the OAM).

In the following the QW2-pol2 box will always be used as a $J_\mathfrak{R}$ operator. The polarization of the state after this box is further modified with an additional quarter-wave plate QW3. This latter modification does not change the charge of the light state but generally, an elliptical state is obtained from an angle $\theta$ between the QW3 slow axis and the x-horizontal axis. The resulting change of the beam spin will have drastic influence on the final OAM after BC2 as it is shown in the following.

We use as input field a Gaussian beam: $\left[\widehat{G} = \widehat{G}(0)\exp(-\frac{w_0^2(k_x^2+k_y^2)}{4})\right]_{xy}$. With the help of Eq. (4) for BC1 (KGW), the field state obtained after the $J_\Re$ projector and QW3 can be summarized as:

$$\begin{bmatrix}1\\0\end{bmatrix}\widehat{G} \rightarrow BC1 \rightarrow \begin{bmatrix}\widehat{\mathfrak{L}}_1^{(0)}\\\widehat{\mathfrak{R}}_1^{(1)}\end{bmatrix}\widehat{G} \rightarrow J_\Re \rightarrow \frac{\widehat{\mathfrak{R}}_1^{(1)}}{\sqrt{2}}\begin{bmatrix}1\\1\end{bmatrix}\widehat{G} \rightarrow QW3(\theta) \rightarrow \frac{\widehat{\mathfrak{R}}_1^{(1)}}{\sqrt{2}}\begin{bmatrix}1+ie^{2i\theta}\\1+ie^{-2i\theta}\end{bmatrix}\widehat{G} \quad (9)$$

This is a charge 1 state whatever the $\theta$ angle. The $\theta$ angle determines the field spin which is the sum (from EQ. (6)) of a LC component $S_L(\theta)$ and a RC one $S_R(\theta)$:

$$S_L(\theta) = \frac{1+\sin(2\theta)}{2} \quad (10)$$

$$S_R(\theta) = -\frac{1-\sin(2\theta)}{2} \quad (11)$$

Applying Eq. (4) again the field after the BZBO=BC2 crystal was calculated with the following step:

$$\rightarrow BC2 \rightarrow \frac{1}{2}\begin{bmatrix}(1+ie^{-2i\theta})\widehat{\mathfrak{L}}_2^{(0)}\widehat{\mathfrak{R}}_1^{(1)} + (1+ie^{2i\theta})\widehat{\mathfrak{L}}_2^{(-1)}\widehat{\mathfrak{R}}_1^{(1)}\\(1+ie^{-2i\theta})\widehat{\mathfrak{R}}_2^{(1)}\widehat{\mathfrak{R}}_1^{(1)} + (1+ie^{2i\theta})\widehat{\mathfrak{R}}_2^{(0)}\widehat{\mathfrak{R}}_1^{(1)}\end{bmatrix}\widehat{G} \quad (12)$$

As we can see, at the exit of the BC2 crystal, the light state is a coherent superposition of several $\widehat{L}_z$ eigenstates with integer OAM. We have selected some of them playing with the optical elements located after BC2.

The state obtained from the QW4-Pol3 box used as a $J_\Re$ operator is:

$$|J\_R> = \frac{1}{2\sqrt{2}}\{(1+ie^{-2i\theta})\widehat{\mathfrak{R}}_2^{(1)}\widehat{\mathfrak{R}}_1^{(1)} + (1+ie^{2i\theta})\widehat{\mathfrak{R}}_2^{(0)}\widehat{\mathfrak{R}}_1^{(1)}\}\widehat{G}\begin{bmatrix}1\\1\end{bmatrix} \quad (13)$$

while from the box used as a $J_\mathfrak{L}$ operator it is:

$$|J\_L> = \frac{1}{2\sqrt{2}}\{(1+ie^{-2i\theta})\widehat{\mathfrak{L}}_2^{(0)}\widehat{\mathfrak{R}}_1^{(1)} + (1+ie^{2i\theta})\widehat{\mathfrak{L}}_2^{(-1)}\widehat{\mathfrak{R}}_1^{(1)}\}\widehat{G}\begin{bmatrix}1\\1\end{bmatrix} \quad (14)$$

The state obtained by fixing $\theta = 0°$ and $\alpha_4$ continuously varying is:

$$|\alpha_4> = \frac{1+i}{4\sqrt{2}} \widehat{\mathfrak{R}}_1^{(1)} \left\{ \left(1 + ie^{2i\alpha_4}\right)\left(\widehat{\mathfrak{L}}_2^{(0)} + \widehat{\mathfrak{L}}_2^{(-1)}\right) + \left(1 + ie^{-2i\alpha_4}\right)\left(\widehat{\mathfrak{R}}_2^{(0)} + \widehat{\mathfrak{R}}_2^{(1)}\right) \right\} \widehat{G} \begin{bmatrix} 1 \\ 1 \end{bmatrix} \quad (15)$$

## 4. Results

The light states versus their OAM described by Eq. (13), (14) and (15) are studied in the following three subsections. Efforts were made to study the field inside the dark regions of the field intensity where the phase vortices are located. The light intensities and specially the dark region locations, measured by the CCD camera and calculated theoretically, are represented in Fig. 2. The contrast of the interference fringes in the dark regions was improved sometimes to the detriment of the bright regions.

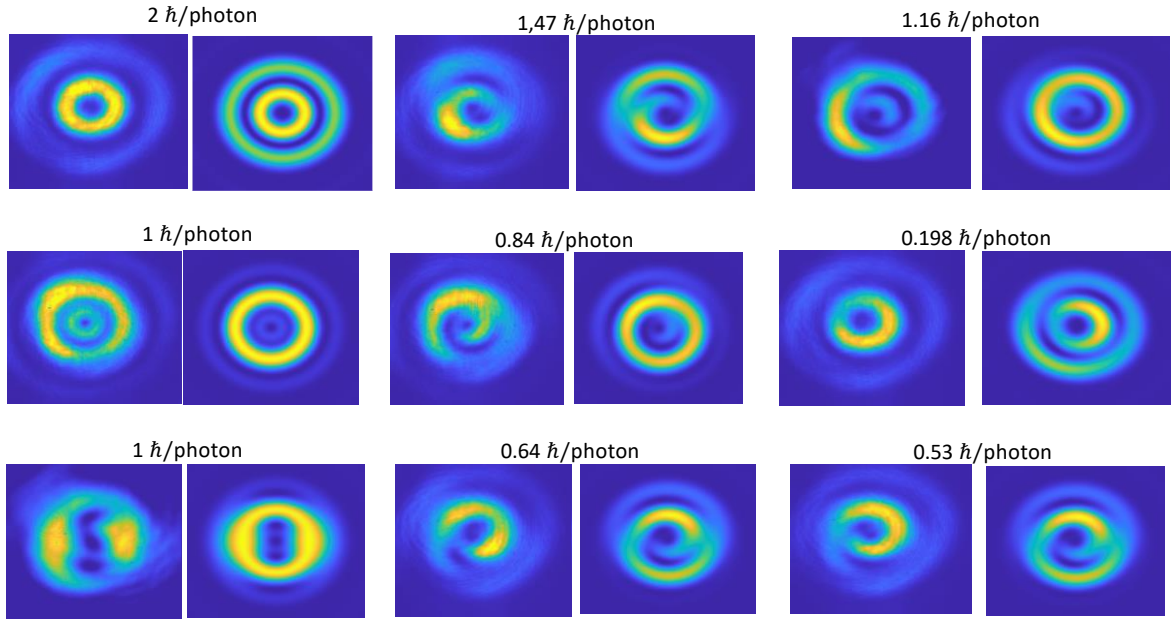

Fig. 2 Intensity patterns in the far field for various OAM. For each OAM the left picture is experimental, the right one id theoretical.

### 4.1 Fractional OAM in the |1, 2] $\hbar$/photon interval

Eq. (13) shows that after the $J_{\mathfrak{R}}$ operator, the light state $|J\_R >$ is a coherent superposition of the two integer OAM states:

$$|1a\rangle = \widehat{\mathfrak{R}}_2^{(0)}\widehat{\mathfrak{R}}_1^{(1)}\widehat{G}\begin{bmatrix}1\\1\end{bmatrix}$$

$$|2\rangle = \widehat{\mathfrak{R}}_2^{(1)}\widehat{\mathfrak{R}}_1^{(1)}\widehat{G}\begin{bmatrix}1\\1\end{bmatrix}$$

whose charge is respectively 1 and 2. So the superposition has generally a fractional OAM in the |1, 2] $\hbar$/photon interval, given by the following formula obtained from inserting Eq. (13) into Eq. (5):

$$l_z = \frac{2(1+\sin(2\theta))\langle 2|2\rangle + (1-\sin(2\theta))\langle 1a|1a\rangle}{(1+\sin(2\theta))\langle 2|2\rangle + (1-\sin(2\theta))\langle 1a|1a\rangle} \quad \hbar/\text{photon} \quad (15)$$

with the numerical values $\langle 1a|1a\rangle = 2.83\ 10^{-3}\ a.u.$, $\langle 2|2\rangle = 2.49\ 10^{-3} a.u.$, where we have benefited from the states orthogonality calculated to be $\langle 2|1a\rangle \cong 10^{-10} a.u.$ (which is much weaker than $\langle 2|2\rangle$ and $\langle 1a|1a\rangle$).

Combining now Eq. (15) and (10) we can exhibit the spin-orbit coupling with the LC component of the incident state in the BC2 crystal responsible for the $l_z$ increasing from 1 $\hbar$/photon:

$$l_z(\theta) = 1 + \frac{2\langle 2|2\rangle}{(1+\sin(2\theta))\langle 2|2\rangle + (1-\sin(2\theta))\langle 1a|1a\rangle} S_L(\theta) \quad (16)$$

where the coefficient of $S_L(\theta)$ is a fraction found numerically close to 1 (it should be exactly 1 if the eigenstates had the same modules or if $\theta = \pi/4$).

In the extreme case $\theta = \pi/4$ (intensity in Fig. 2 (a) and (a')), the OAM is 2 $\hbar$/photon and the interference pattern with the reference beam exhibits at its centre a vortex characterized by two supplementary bright fringes located in the lower part of a circle centred on the vortex (red arrow inside Fig. 3 (a)). This corresponds to $4\pi$ rad phase variation along a path around it as the theoretical Fig. 3 (a') shows. The two -π/+π discontinuous phase lines near the centre in Fig. 3 (a') reveal the fact that the phase-front is helicoidal. As soon as the $\theta$ angle decreases towards 0, this charge-2 vortex splits into two charge +1 vortices and $l_z$ becomes fractional. This is shown in the case $\theta = 0$ (intensity in Fig. 2 (b) and (b'), red arrows near the middle of Fig. 3 (b) and middle of Fig. 3 (b')). The $l_z$ value is 1.47 $\hbar$/photon, that is to say very close to the 1.5 $\hbar$/photon half-integer value. Simultaneously a new other charge -1 vortex appears (upper red arrow in Fig. (3b)) which is linked by a bright fringe to one previous charge +1 vortex (this link is underlined by the two linked red arrows). The theoretical Fig. 3 (b') shows also the link between these two opposite signs vortices. Simultaneously another charge +1 vortex appears (bottom red arrow in Fig. 3 (b) and Fig. 3 (b')). All these vortices and their links evolve continuously up to a value of the $\theta$ angle slightly negative. Then a rearrangement occurs, shown in Fig. (3c) and (3c') for $\theta = -0.111\pi$ ($l_z = 1.16\ \hbar$/photon, intensity in Fig. 2 (c) and (c')) and a new linked between two opposite charge vortices appears (linked red arrows). These two latter vortices and their link disappear in a continuous evolution (not shown in Fig. 3) when the $\theta$ angle decreases towards $\theta = -\pi/4$ ($l_z = 1\ \hbar$/photon).

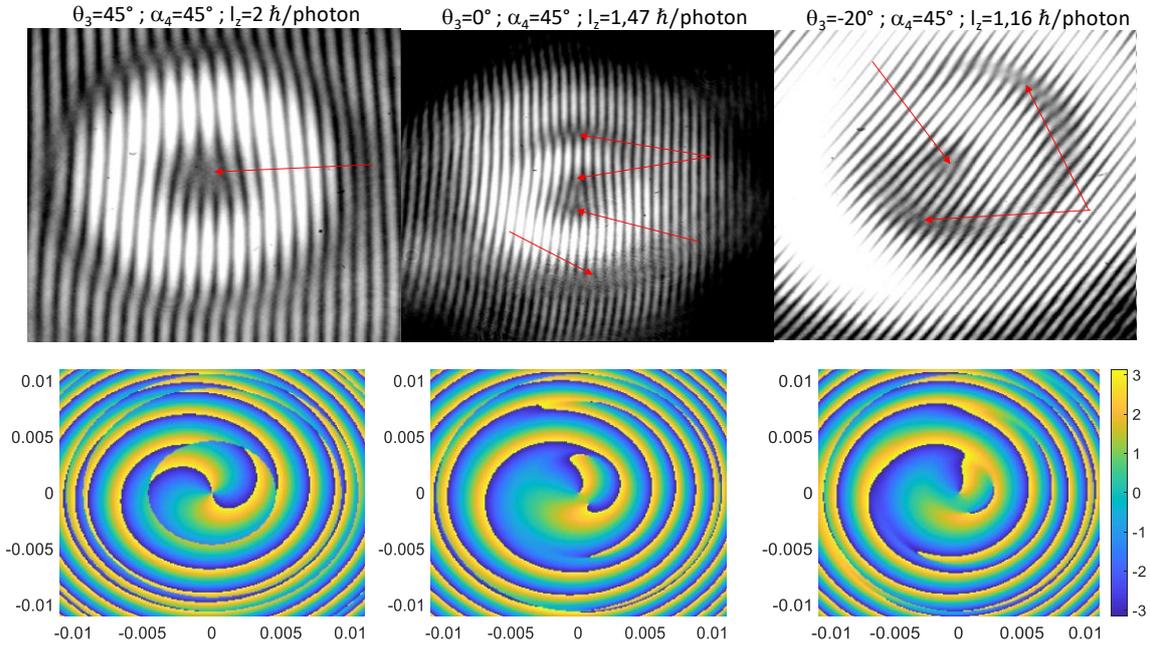

Fig. 3 Interferogram and phase pattern of the $|J\_R>$ state with various OAM.

### 4.2 Fractional OAM in the |0, 1] $\hbar$/photon interval

Eq. (14) shows that after the $J_\mathfrak{L}$ operator, the light state $|J\_L>$ is a coherent superposition of the two integer OAM states:

$$|0> = \widehat{\mathfrak{L}}_2^{(-1)} \widehat{\mathfrak{R}}_1^{(1)}\ \widehat{G} \begin{bmatrix} 1 \\ 1 \end{bmatrix}$$

$$|1b> = \widehat{\mathfrak{L}}_2^{(0)} \widehat{\mathfrak{R}}_1^{(1)} \widehat{G} \begin{bmatrix} 1 \\ 1 \end{bmatrix}$$

whose charge is respectively 0 and 1. The resulting fractional OAM in the |0, 1] $\hbar$/photon interval, given by the following formula:

$$l_z = \frac{(1+\sin(2\theta))<1b|1b>}{(1+\sin(2\theta))<1b|1b>+(1-\sin(2\theta))<0|0>} \; \hbar/\text{photon} \quad (17)$$

with the numerical values $<1b|1b> = 2.83 \; 10^{-3} \; a.u.$, $<0|0> = 2.49 \; 10^{-3} a.u.$, and where we have benefited from the states orthogonality calculated to be $<0|1b> \cong 10^{-10} a.u.$.

Combining now Eq. (17) and (11) we can exhibit the spin-orbit coupling in the BC2 crystal with the RC component of the incident state responsible for the $l_z$ decreasing from 1 $\hbar$/photon:

$$l_z(\theta) = 1 + \frac{2<0|0>}{(1+\sin(2\theta))<1b|1b>+(1-\sin(2\theta))<0|0>} S_R(\theta) \quad (18)$$

where the coefficient of $S_R(\theta)$ is a fraction found numerically close to 1

In the case $\theta = \pi/4$ (intensity in Fig. 2 (d) and (d')), the OAM is 1 $\hbar$/photon and the interference pattern with the reference beam exhibits at its centre a vortex characterized by one supplementary bright fringe located in the lower part of a circle centred on the vortex (red arrow inside Fig. 4 (a)). This corresponds to $2\pi$ rad phase variation along a path around it as the theoretical Fig. 4 (a') shows. The $-\pi/+\pi$ discontinuous phase line near the centre in Fig. 4 (a') reveal the fact that the phase-front is helicoidal. As soon as the $\theta$ angle decreases towards 0, this $-\pi/+\pi$ discontinuous phase line breaks and two new vortices with opposite charges are born near the centre. This is shown in Fig. 4 (b) and (b') (red arrows) corresponding to $\theta = +0.111\pi$ ($l_z = 0.84$ $\hbar$/photon, intensity shown in Fig. 2 (e) and (e')). The new -1 charge vortex is linked with + 1 charge original vortex (pointed out by the two linked red arrows). Then a continuous evolution of this structure is observed when $\theta$ and $l_z$ decrease. This is shown in Fig. 5 (c) and (c') for $\theta = 0$ and $l_z = 0.53$ $\hbar$/photon (intensity represented in Fig. 2 (i) and (i')). The values $\theta = -0.111\pi$ and $l_z = 0.113$ $\hbar$/photon lead to Fig. 4 (c) and (c') (here the experimental resolution is not sufficient to distinguish the two opposite charge vortices in the centre), the intensity being shown in Fig. 2 (f) and (f')); all the vortices disappear for $\theta = -\pi/4$ and $l_z = 0$ $\hbar$/photon (this case is not shown).

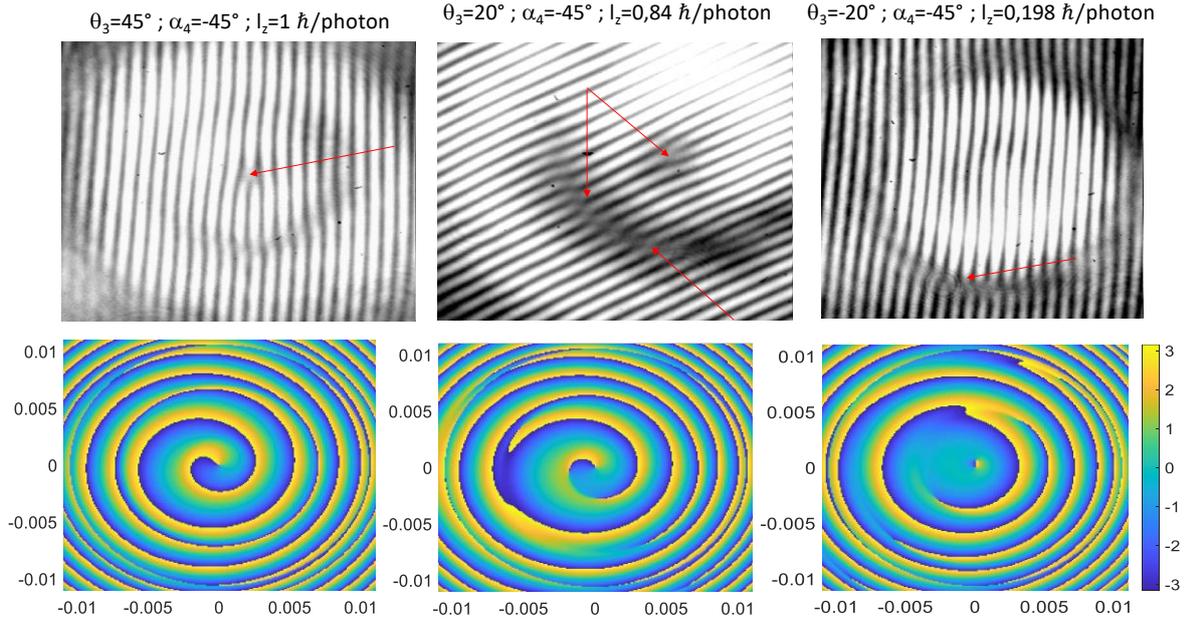

Fig. 4 Interferogram and phase pattern of the $|J\_L>$ state with various OAM.

### 4.3 Composite states with total OAM around 1 $\hbar$/photon

Eq. (15) shows that the $|\alpha_4>$ state is a superposition of the $|0>, |1a>, |1b>$ and $|2>$ states. The resulting fractional OAM is given by the following formula:

$$l_z = \frac{2(1+\sin(2\alpha_4))<2|2>+(1+\sin(2\alpha_4))<1a|1a>+(1-\sin(2\alpha_4))<1b|1b>}{(1+\sin(2\alpha_4))<2|2>+(1+\sin(2\alpha_4))<1a|1a>+(1-\sin(2\alpha_4))<1b|1b>+(1-\sin(2\alpha_4))<0|0>}$$

$\hbar$/photon

(19)

where we have used $<1a|1b>\cong 10^{-5} a.u., <2|1b>\cong 10^{-10} a.u., <0|1a>\cong 10^{-11}, <0|2>\cong 10^{-11} a.u.$

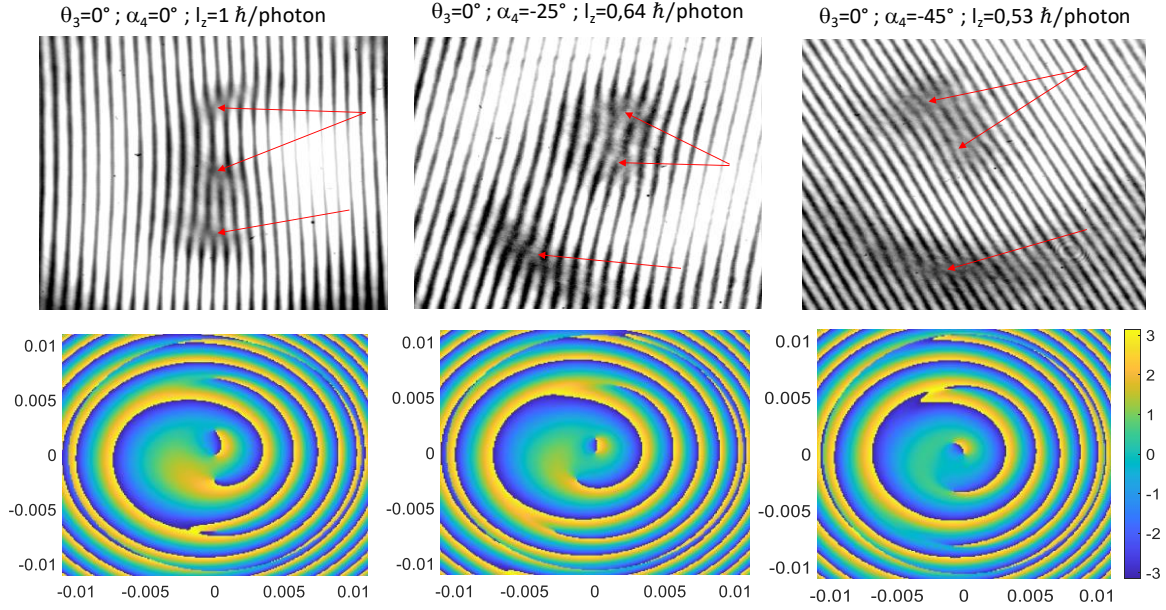

Fig. 5 Interferogram and phase pattern of the $|\alpha_4>$ state with various OAM.

Let us start with the state $|\alpha_4 = 0>$ obtained with the $\alpha_4 = 0$ angle. We have $l_z = 1\ \hbar/$photon but despite that this value is integer, the state intensity (Fig. 2 (i) and i')) has not a circular symmetry unlike the $|1b>$ eigen-state (Fig. 2 (d) and d')). This is of course because this $|\alpha_4 = 0>$ state is not a pure state of OAM but is a coherent superposition of four eigen-modes. This intensity vanishes in three points where we expect vortices. This is what we found in Fig 5 (a) and (a'). Two opposite charge vortices are linked by a -π/+π discontinuous phase line (linked red arrows). When decreasing the $\alpha_4$ angle, the intensity, interference pattern and phase evolves continuously. This is shown for the state $|\alpha_4 = -0.139\pi>$ ($l_z = 0.642\ \hbar/$photon, Fig 2 (j) and (j'), Fig 5 (b) and (b')), and for the state $|\alpha_4 = -\pi/4>$ ($l_z = 0.53\ \hbar/$photon, Fig 2 (k) and (k'), Fig 5 (c) and (c')).

### 4.4 Discussion of the results

We have shown that a pair of vortices of integer charge of the same sign (+1) is born when lowering from $l_z = 2\ \hbar/$photon the OAM of the light state. We have shown that the continuous evolution of the phase pattern going with the $l_z$ evolution is sometimes interrupted by the breaking of a -π/+π discontinuous phase line linking two existing opposite charge vortices. In this case a new pair of vortices with opposite charges is born. If we inspect the case of the phase pattern evolution around half-integer OAM, that is to say around 1.5 (Fig. 3 (b) and (b')) and 0.5 $\hbar$/photon (Fig. 5 (c) and (c')), we observe a continuous evolution and not a specific pattern or a sudden transition. This is in contrast with the change of vortex topology reported in previous works concerning the phase pattern of fields created by spiral phase plates with fractional step height or spatial light modulator. In these cases, for half-integer phase steps, a chain of additional vortices with alternating charge is observed experimentally and theoretically in the region of low intensity [15-17].

## 5. Conclusions

The mechanisms leading to the evolution of light sates versus their fractional OAM from a CP Gaussian beam with no OAM have been pointed out in the cascaded CD process. Experimental results are provided by a non-degenerate cascade of two different KGW and BZBO biaxial crystals. The 0→1→2 $\hbar$/photon increasing of the OAM in passing through fractional values was operated with the help of the spin-orbit coupling in the BZBO crystal. The phase of the state light and its vortices were visualized by interference patterns with a reference beam. The evolution of the fractional OAM value is accompanied by a continuous evolution of pairs of vortices with opposite signs and linked by a -π/+π discontinuous phase line. The phase pattern evolution around half-integer OAM, that is to say around 1.5 and 0.5 $\hbar$/photon, is observed to be continuous and not with a specific or a sudden behavour. In other cases, the evolution can be interrupted by the breaking of a -π/+π discontinuous phase line and a new pair of vortices with opposite charges is born. All the experimental patterns (intensity, phase, OAM) are quite well described by a full numerical mode exhibited in the Fourier space.

### Acknowledgments

We thank A. Majchrowski and E. Michalski for providing the BZBO sample.

**Figure captions**

1. Experimental set-up allowing the OAM study in the far field. The CD-field in position 1 is transported without modification in position 2 by the L2-L3 afocal telescope.
2. Intensity patterns in the far field for various OAM. For each OAM the left picture is experimental, the right one id theoretical.
3. Interferogram and phase pattern of the $|J\_R>$ state with various OAM.
4. Interferogram and phase pattern of the $|J\_L>$ state with various OAM.
5. Interferogram and phase pattern of the $|\alpha_4>$ state with various OAM.